# Estimation of Solar Spectral Irradiance Using Meteorological Data and Analysis of Optimal Conditions for Solar Power Generation

Jeonggyu Hwang[1, *]

[1]*Department of Semiconductor Engineering,*
*Gachon University, Seongnam-si, Gyeonggi-do, South Korea 131200*

This study proposes an approximate model to estimate the solar radiation spectrum intensity in Seoul, Republic of Korea, for the year 2024, aiming to analyze optimal conditions related to energy generation. Since the solar radiation spectrum varies with atmospheric conditions, accurately predicting it typically requires complex spectral radiation models. However, such models entail high computational costs, hindering real-time application. To address this, this study introduces a simplified approximation model using only direct normal irradiance (DNI) among real-time meteorological elements, employing linear scaling of the standard spectrum (ASTM G-173). This model first estimates DNI using global horizontal irradiance (GHI) and solar position information (such as zenith angle), then linearly adjusts the standard spectrum to compute real-time spectrum intensity. The model approximates realistic DNI values by correcting various meteorological parameters, including zenith angle, cloud cover, and visibility. The analysis shows that GHI exhibits stable seasonal patterns, peaking in summer and minimizing in winter. In contrast, DNI demonstrates significant temporal variability and frequent abnormal peaks (e.g., exceeding 9,000 W/m²), highlighting the importance of data refinement and anomaly detection in predicting energy generation. In conclusion, GHI is suitable for general photovoltaic analyses, whereas DNI is crucial for direct-beam sensitive systems like concentrated solar power (CSP), requiring meticulous data quality management. Future research should focus on identifying the causes of DNI anomalies and developing real-time quality control algorithms.

The solar radiation spectrum serves as a critical input parameter across various applications, including solar power generation, thermal radiation modeling, and climate analysis[1,2]. In reality, solar radiation in the atmosphere undergoes wavelength-dependent attenuation due to various atmospheric constituents such as water vapor, ozone, and aerosols. Accurately predicting these spectral changes necessitates the use of spectral radiation models.

However, such models have practical limitations due to their high computational costs[3]. To address this issue, the present study proposes an approximate model that linearly adjusts the spectral intensity based solely on Direct Normal Irradiance (DNI) measured under real-time weather conditions, referencing the standard solar spectrum.

This study aims to estimate spectral intensities in Seoul, Republic of Korea, throughout the year 2024 using the proposed approximate model. By analyzing the estimated values, the goal is to identify factors influencing energy output and determine optimal conditions for maximizing solar power generation.

## Methodology
### Nature of Standard Spectrum

The ASTM (American Society for Testing and Materials) G-173 model presented in Fig. 1. is a highly reputable solar spectrum dataset provided by the U.S. National Renewable Energy Laboratory (NREL), widely recognized in the renewable energy field. This dataset is composed of calculated values of spectral irradiance—representing solar radiation energy as a function of wavelength—organized by wavelength.

| | |
|---|---|
| Air Mass | AM 1.5 |
| Solar Zenith Angle | ≈ 48.2° |
| Precipitable Water Vapor | 1.42 cm |
| Ozone Concentration | 0.34 atm-cm |
| Aerosol Optical Depth (AOD) | 0.084 |

Table 1. Summary of the Basic Conditions of ASTM G-173 Spectrum

The dataset is based on geographical coordinates of Golden, Colorado, USA (Latitude: 39.74°N, Longitude: 105.18°W, Altitude: 1,828 m), and is derived through radiative transfer simulations utilizing Beer–Lambert's law under the atmospheric conditions outlined in Table 1.

This dataset represents a static solar spectrum that already incorporates wavelength-dependent absorption characteristics, defined as follows:

$$I_{\text{std}}(\lambda): \lambda \in [280, 2500] \text{ nm} \tag{1}$$

---

[*] h5638880@gachon.ac.kr

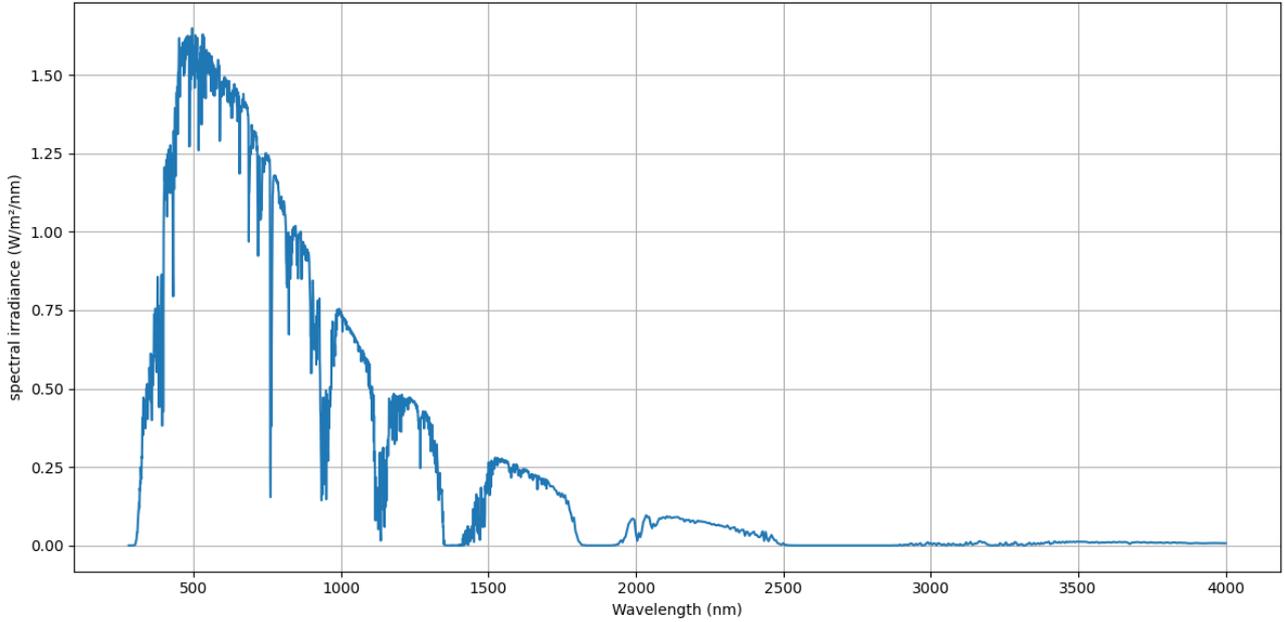

Figure 1. Detailed spectral irradiance for light wavelengths.

When normalized as a wavelength distribution function, it can be expressed as follows:

$$f(\lambda) = \frac{I_{\text{std}}}{\int_{\lambda_0}^{\lambda_1} I_{\text{std}}(\lambda)\, d\lambda} \quad (2)$$

Here, $f(\lambda)$ is a dimensionless function representing only the energy distribution, normalized to have an area of 1

**Nature of Standard Spectrum**

The Direct Normal Irradiance (DNI), which can be obtained hourly from meteorological observation data, is defined as follows:

$$E = \int_{\lambda_0}^{\lambda_1} I(\lambda)\, d\lambda \quad (3)$$

Based on the total direct energy $E$, the overall solar spectrum is approximated through linear scaling, as shown below:

$$I(\lambda) = f(\lambda) \cdot E \quad (4)$$
$$E = \text{DNI}\ [W/m^2] \quad (5)$$

This approximation relies on the following assumptions:
- Assumption 1:
 Wavelength-dependent absorption characteristics do not vary over time.
- Assumption 2:
 Primary atmospheric changes affect only the total DNI.
- Assumption 3:
 The spectral shape per wavelength remains fixed at $f(\lambda)$, calculated under standard conditions.

The core strength of this model is that it maintains the spectral shape while still accounting for hourly variations in solar intensity.

**Physical Interpretation**

This method employs an approximation of Beer–Lambert's law, which can be expressed as follows:

$$I(\lambda) = I_0(\lambda) \cdot e^{-\tau(\lambda)} \quad (6)$$

Here, $\tau(\lambda)$ represents the atmospheric optical depth, a wavelength-dependent variable influenced by water vapor, ozone, aerosols, and scattering effects.

The standard spectrum $I_{\text{std}}(\lambda)$ can thus be viewed as the result of calculating this optical depth under a specific set of conditions.

Consequently, the linear scaling model described earlier takes the following form:

$$I(\lambda) \approx I_{\text{std}}(\lambda) \cdot \frac{E_{\text{real}}}{E_{\text{std}}} \quad (7)$$

This indicates that only the total irradiance is updated, while the wavelength-dependent attenuation remains unchanged.

**DNI Estimation from GHI**

Using trigonometric functions, the following relationship for directional solar irradiance and projected surface area can be derived:
$$GHI = DHI + DNI \cdot \cos(\theta_z) \tag{8}$$

Here, GHI is the Global Horizontal Irradiance—representing the total irradiance reaching a horizontal surface. DNI (Direct Normal Irradiance) is the direct irradiance perpendicular to a surface facing the sun, while $\theta_z$ is the solar zenith angle, and $\cos(\theta_z)$ is the horizontal projection coefficient depending on solar altitude.

In this model, the following approximation using only GHI is adopted[4]:
$$DNI \approx \frac{GHI}{\cos(\theta_z)} \tag{9}$$

Overall, this model approximates DNI based solely on GHI and solar positional data (zenith angle). It linearly scales the normalized standard solar spectrum using the calculated DNI, providing a simplified method to estimate the hourly solar radiation spectrum. Therefore, the primary objective of this research is to find and analyze optimal values centered around DNI.

This method allows computations using only real-time meteorological data, significantly reducing computational load compared to classical radiative transfer models, while reliably capturing the total amount and variability of solar radiation.

## Results
### Model Factor

In this study, a physics-based correction model was applied to estimate DNI from the observed Global Horizontal Irradiance (GHI), using meteorological data provided by the Korea Meteorological Administration for the year 2024. This estimation formula considers meteorological parameters such as solar zenith angle, cloud cover, and visibility to approximate the actual direct irradiance component.

The Direct Normal Irradiance can theoretically be approximated from GHI using the solar zenith angle as follows:
$$DNI_0 = \frac{GHI}{\cos(\theta_z)} \tag{10}$$

where $\theta_z$ is the solar zenith angle. However, when $\cos(\theta_z) \leq 0$ or $GHI \leq 0$, direct irradiance does not physically exist, and thus DNI is treated as zero.

To reflect solar radiation attenuation due to clouds, the total cloud cover was incorporated using the following correction formula for cloud attenuation coefficient[5,6]:
$$DNI_1 = DNI_0 \cdot (1 - 0.75 \cdot cloud\_factor) \tag{11}$$
$$cloud\_factor = \frac{cloud\ cover}{10} \tag{12}$$

Here, the coefficient 0.75 is an empirically derived value indicating approximately 75% attenuation under fully overcast conditions (cloud cover = 10).

Visibility, an indicator of atmospheric clarity, significantly affects solar radiation; lower visibility corresponds to higher radiation attenuation. Visibility effects were incorporated using the following atmospheric transmittance model:
$$DNI_{final} = DNI_1 \cdot \tau_{visibility} \tag{13}$$
$$\tau_{visibility} = \exp\left(-\frac{1}{visibility_{km} + \varepsilon}\right) \tag{14}$$
$$visibility_{km} = \frac{Visibility(10m) \cdot 10}{1000} \tag{15}$$

Here, $\varepsilon = 10^{-6}$ is a minimal threshold value to prevent division by zero. This model ensures that shorter visibility leads to a reduced $\tau_{visibility}$ thus intensifying attenuation of direct irradiance.

Through this correction process, it was possible to estimate DNI while comprehensively reflecting various meteorological factors such as solar zenith angle, cloud cover, and visibility.

### Regression Models

In this study, GHI, $\cos(\theta_z)$, total cloud cover, and visibility were set as independent variables, while DNI was used as the dependent variable in building multiple regression models. The experimental models include: (1) a linear regression model (Polynomial + Ridge) with a second-order polynomial transformation, (2) Decision Tree, (3) Random Forest, (4) Gradient Boosting, and (5) a Multilayer Perceptron (Neural Network, MLP). The goal of this research is to compare the predictive performance of each model and ultimately identify the factors influencing solar power generation, in order to estimate power output.

In this study, four variables were used as features: total cloud cover, visibility, GHI, and $\cos(\theta_z)$. The target variable was DNI. The entire dataset was split into training data (80%) and test data (20%) so that every model would be trained and evaluated under identical conditions.

Overview of the Regression Models
1. Polynomial + Ridge
To address the simplicity of a linear regression model, a second-order polynomial transformation was applied. For each of GHI, $\cos(\theta_z)$, cloud cover, and visibility, second-order and cross terms were generated. Ridge regression (L2 regularization) was then employed to mitigate overfitting.
2. Decision Tree
The decision tree method has the advantage of capturing nonlinear relationships and interactions among variables. In this study, parameters such as maximum depth and minimum samples required for splitting were preset.

3. Random Forest

Random Forest involves training multiple decision trees and averaging their predictions to generate a final output. It was employed to reduce the bias of a single decision tree and enhance predictive stability. Parameters such as the number of trees (n_estimators) and maximum depth (max_depth) were determined through an experimental range.

4. Gradient Boosting

The Gradient Boosting regression model iteratively trains base trees, adjusting for the errors of the previous stage. This model's performance can vary significantly depending on combinations of hyperparameters such as learning_rate, n_estimators, and max_depth.

5. Neural Network (MLP)

A deep neural network in the form of a Multilayer Perceptron was introduced to learn nonlinearity. The input layer was set up to receive the four variables above, and two hidden layers (each with 64 nodes) were included. The Adam optimizer was used for weight optimization, and mean squared error (MSE) served as the loss function.

Model Evaluation Metrics and Procedure

The predictive performance of each model was compared using the coefficient of determination ($R^2$) and the mean squared error (MSE). The $R^2$ indicates how well the model explains the actual data, and a negative value means that the model performs worse than simply predicting the mean of the dataset. A larger mean squared error indicates a greater discrepancy between predicted and actual values.

Comparison of Model Performance

1. Polynomial + Ridge

When a polynomial transformation was applied, followed by Ridge regularization, the $R^2$ was approximately 0.11. This suggests a moderate improvement in the model's ability to explain the data compared to a simple linear model. The mean squared error was about $2.1 \times 10^6$, showing relatively decent predictive performance compared to other models.

2. Decision Tree

Despite limitations on maximum depth and minimum sample splits, the decision tree exhibited an $R^2$ of around -2.60. This implies that the model may have overfitted certain segments or that the tree could not find sufficient splitting criteria due to the nature of the variable distributions. Its mean squared error was also higher than that of other models, indicating that further parameter tuning is critical.

3. Random Forest

The $R^2$ score was about -0.32, which was negative. Although random forests typically provide more stable predictions than a single tree, it appears that the chosen hyperparameters were not well-suited to the data characteristics, or the model structure could not fully avoid overfitting.

4. Gradient Boosting

The $R^2$ score was approximately -0.31, which is similar to that of the random forest. Although boosting methods often yield high performance, the current dataset or initial settings might not have been sufficiently optimized to show a clear performance gain.

5. Neural Network (MLP)

Even with a more complex structure than a single-layer perceptron, the $R^2$ was about -0.03, which is relatively low. This could be due to initial hyperparameter settings (number of hidden layers, number of neurons, learning rate, etc.) remaining at baseline, or insufficient use of regularization techniques (such as dropout or batch normalization) to counter overfitting.

Analysis of Factors Affecting Model Performance

Among the various models, only the polynomial-transformed Ridge regression achieved a positive $R^2$ value. As illustrated in Fig. 2, suggesting that a simple linear model alone cannot adequately explain the data, but that nonlinear terms and regularization can be somewhat effective. It is also possible that variables like cloud cover, visibility, and $\cos(\theta_z)$ do not fully capture the factors determining DNI. For instance, incorporating humidity, aerosol concentration, and seasonal patterns might yield better outcomes in tree-based ensemble models or neural networks. Moreover, this study did not specifically account for time-series characteristics, so abrupt changes in solar altitude during certain times of the day or seasonal variability may not have been fully reflected in the models.

**Result Analysis**

The results of the time-series analysis for daily average GHI and DNI from January to December 2024 are shown in Fig. 3. The graph intuitively illustrates how GHI and DNI vary seasonally, with GHI plotted against the left Y-axis (units: W/m²) and DNI against the right Y-axis (units: W/m²).

GHI exhibited a clear increasing trend from spring to early summer (March to June), reaching a peak of approximately 580 W/m², coinciding with periods of higher solar altitude and increased daylight hours. Conversely, during the winter months (November to January), GHI dropped significantly, averaging below 100 W/m², clearly indicating strong seasonal variability.

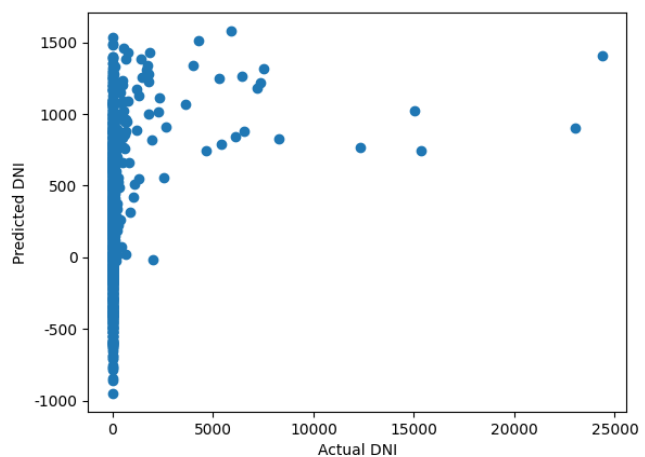

Figure 2. Measured vs. Predicted DNI Scatter Plot

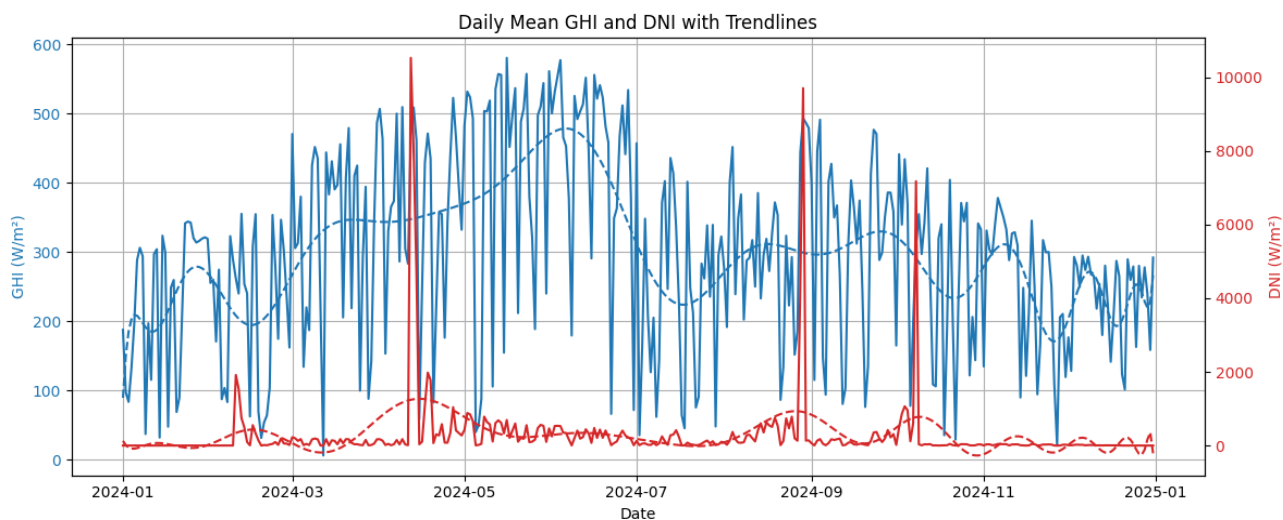

Figure 3. Time series of daily mean GHI and DNI from January to December 2024.

DNI showed considerably greater fluctuations compared to GHI, characterized by sharp peaks at specific points throughout the year. Notably, several extreme peaks exceeding 9,000–10,000 W/m² were observed around April, August, and October. These spikes correspond to highly concentrated direct irradiance and could potentially be intermittent direct sunlight between cloud cover or sensor anomalies. Under normal conditions, DNI typically ranged from 0–500 W/m² on average, generally showing an increasing trend in summer.

Comparing the seasonal patterns of these two metrics, GHI followed a relatively continuous and stable curve, whereas DNI displayed considerable disturbances and short-term variability, allowing complementary interpretations. Additionally, both GHI and DNI recorded high values during summer, indicating that favorable seasons for solar energy generation consistently occur in this particular region.

## Discussion

Analysis of the daily average Global Horizontal Irradiance (GHI) and Direct Normal Irradiance (DNI) for the year 2024 revealed both clear seasonal trends and the presence of abnormal outliers. GHI followed a predictable sinusoidal pattern throughout the year, reflecting variations in solar altitude and atmospheric path length, thus demonstrating stability and reliability as an indicator for evaluating solar energy resources..

In contrast, DNI exhibited several extreme peaks exceeding 9,000 W/m², notably around April, August, and October, in addition to the general seasonal upward trend. Such values are highly unlikely under normal clear-sky conditions and could be attributed to sensor errors, calibration inconsistencies, or extreme weather phenomena like the cloud-edge effect or sudden irradiance recovery. These anomalies are particularly critical for Concentrated Solar Power (CSP) systems, which are highly sensitive to DNI variations, potentially significantly impacting the accuracy of energy yield predictions.

The variability differences between GHI and DNI underscore the importance of selecting appropriate irradiance metrics tailored to system type. While fixed solar systems can reliably utilize GHI-based analysis, solar-tracking or concentrating systems require detailed DNI analysis that accounts for short-term variability and outliers.

Furthermore, the high variability observed in DNI suggests a necessity for incorporating anomaly detection methods and adaptive forecasting algorithms into short-term predictive models or real-time energy management systems.

## Conclusion

This study analyzed daily average Global Horizontal Irradiance (GHI) and Direct Normal Irradiance (DNI) data for the year 2024, leading to the following conclusions. First, GHI exhibited clear seasonal patterns, peaking during summer and decreasing in winter, validating its effectiveness as a resource evaluation metric for typical solar photovoltaic systems. Second, DNI showed significant variability at relatively high temporal resolution, with numerous abnormal spikes observed throughout the year. This highlights the need for accurate outlier handling procedures in energy yield prediction models and Concentrated Solar Power (CSP) system operations. Finally, selecting appropriate irradiance indicators—GHI or DNI—should be aligned with specific solar system characteristics, with particular emphasis on incorporating data quality control and anomaly removal algorithms when utilizing DNI.

Future research should focus on identifying causes of DNI anomalies by conducting supplementary meteorological analyses, integrating satellite-derived cloud data or all-sky camera imagery, and developing real-time quality control algorithms. This study provides fundamental insights and valuable baseline data for designing, forecasting, and optimizing irradiance-based solar energy systems.

| Index | Date/Time 일시 | Solar Radiation (MJ/m2) 일사(MJ/m2) | Total Cloud Amount (10-scale) 전운량(10분위) | Visibility (10m) 시정(10m) |
|---|---|---|---|---|
| 0 | 2024.1.1 00:00:00 | | 0 | 394 |
| 1 | 2024.1.1 01:00:00 | | 0 | 402 |
| 2 | 2024.1.1 02:00:00 | | 6 | 616 |
| 3 | 2024.1.1 03:00:00 | | 1 | 265 |
| 4 | 2024.1.1 04:00:00 | | 6 | 203 |
| 5 | 2024.1.1 05:00:00 | | 8 | 371 |
| 6 | 2024.1.1 06:00:00 | | 8 | 257 |
| … | | | | |
| 8777 | 2024.12.31 17:00:00 | 0.42 | 0 | 3570 |
| 8778 | 2024.12.31 18:00:00 | 0.02 | 0 | 3672 |
| 8779 | 2024.12.31 19:00:00 | | 0 | 3760 |
| 8780 | 2024.12.31 20:00:00 | | 0 | 3645 |
| 8781 | 2024.12.31 21:00:00 | | 0 | 3794 |
| 8782 | 2024.12.31 22:00:00 | | 0 | 3779 |
| 8783 | 2024.12.31 23:00:00 | | 0 | 3412 |

**Extended Data Table 1. | Sample meteorological observations from Seoul, Korea (2024).** A representative excerpt from the hourly automatic synoptic observation system (ASOS) data provided by the Korea Meteorological Administration (KMA). The table includes the first seven and last seven hourly entries recorded in 2024, showing four key variables used in the analysis: (1) Date/Time (일시), (2) Solar Radiation (MJ/m²) (일사), (3) Total Cloud Amount (10-scale) (전운량), and (4) Visibility (10 m) (시정). The full dataset includes 8,784 hourly observations from January 1st to December 31st, 2024. Only selected rows are shown here for brevity; middle rows have been omitted.

a

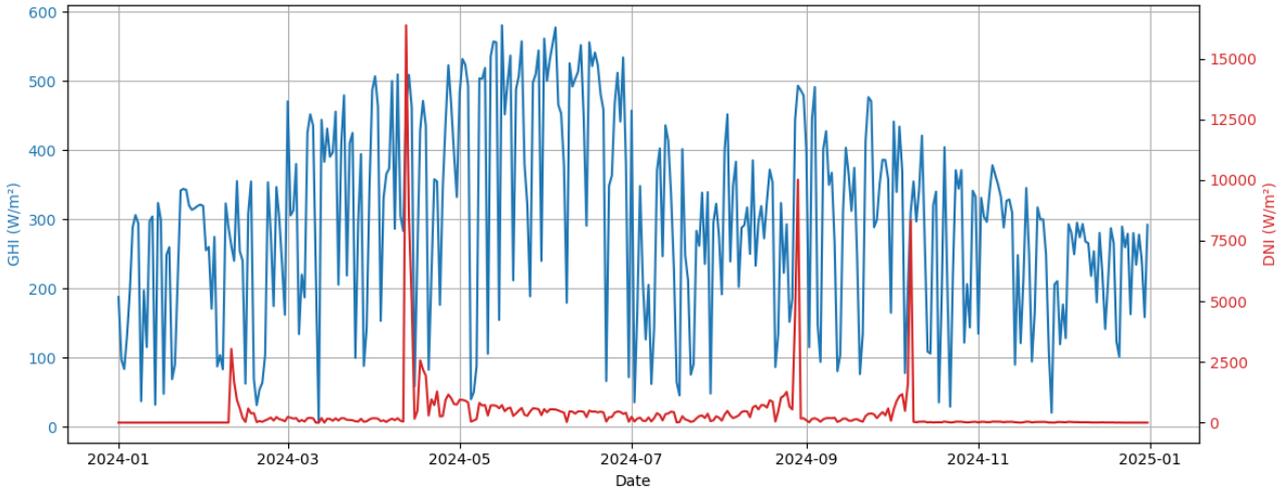

b

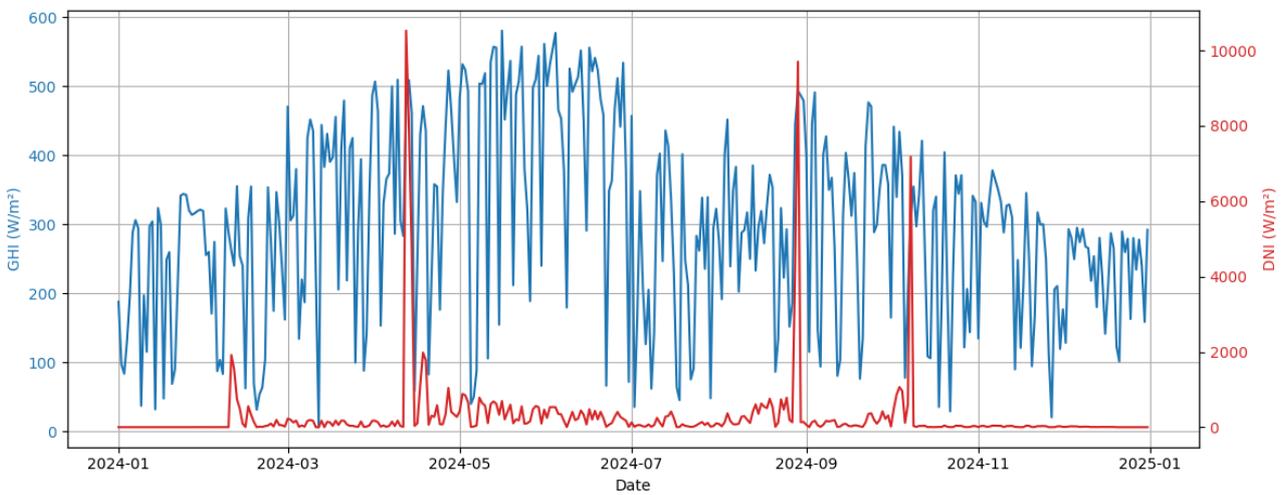

**Extended Data Figure 1. | Comparison of DNI Estimation Before and After Atmospheric Condition Correction. a**, Estimated DNI without atmospheric correction, calculated solely from GHI and solar zenith angle using geometric projection (Equation 9). This approach results in overestimation during cloudy or hazy periods. **b**, DNI corrected using total cloud amount and visibility data. Empirical attenuation factors were applied to reduce DNI values under low-transparency conditions, resulting in a more physically realistic time series.

a

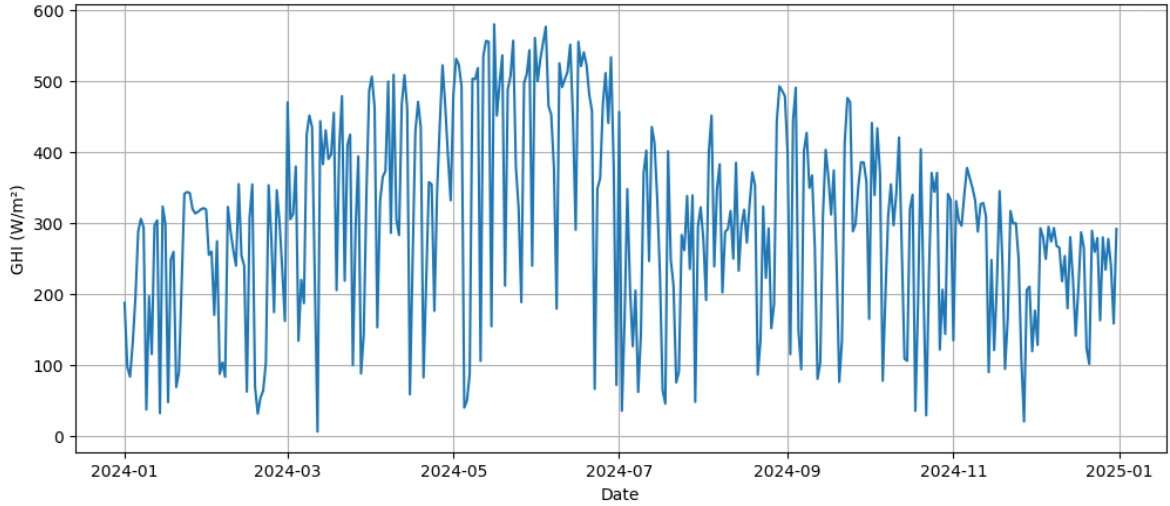

b

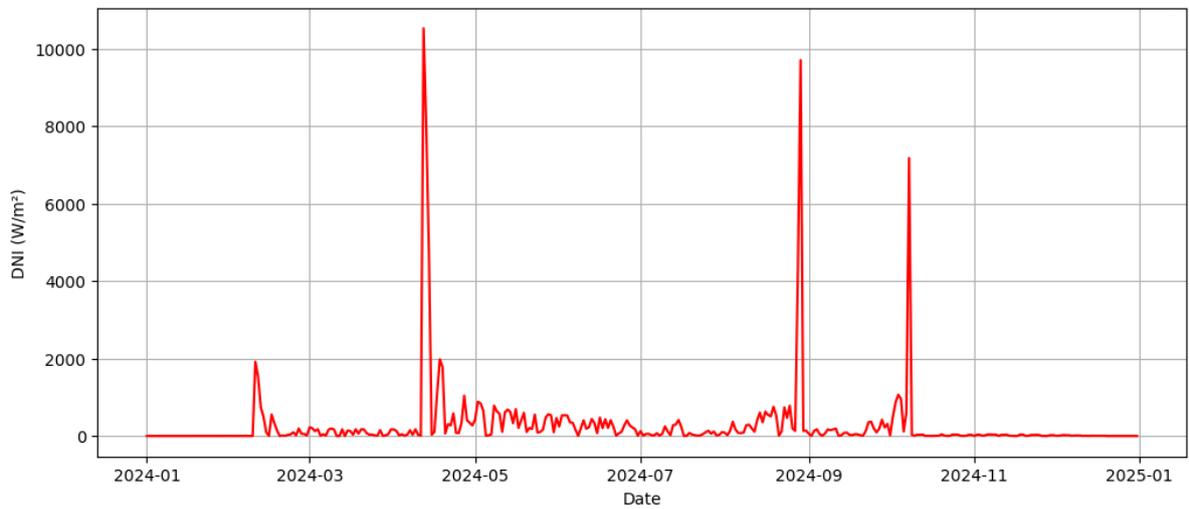

**Extended Data Figure 2. | Separated Time Series of GHI and Estimated DNI for 2024. a**, Daily mean Global Horizontal Irradiance (GHI) measured at the Seoul Observatory throughout 2024. Seasonal variations are clearly observed, with higher irradiance in the spring and summer months. **b**, Direct Normal Irradiance (DNI) estimated using GHI and solar zenith angle, further corrected based on total cloud amount and visibility. The applied empirical attenuation factors significantly suppress unrealistic peaks during cloudy or low-visibility periods, providing a physically consistent representation of surface-level direct irradiance.